\documentclass[runningheads]{llncs}
\usepackage[T1]{fontenc}
\usepackage{graphicx}
\usepackage{amsmath}
\usepackage{amssymb}
\usepackage{array}

\usepackage{booktabs}
\usepackage{multirow}
\usepackage{longtable}
\usepackage{array}
\usepackage{ragged2e}
\usepackage{float}
\usepackage[dvipsnames]{xcolor}

\begin{document}
\raggedbottom
\title{BEAM3R: Beam’s-eye-view architecture with Mamba-3 for implicit dose reconstruction}
\titlerunning{BEAM3R: BEV fast implicit dose reconstruction with Mamba-3}
%
\author{Chen Cheng \inst{1}\orcidID{0009-0007-6612-315X
} \and
Michael Ferraro\inst{1}\orcidID{0009-0002-9211-0321} \and
James Grover \inst{1}\orcidID{0009-0009-6743-7809}
\and
David E J Waddington \inst{1}\orcidID{0000-0002-7017-1556}
\and
Emily Hewson \inst{1}\orcidID{0000-0001-7353-2663}}
\authorrunning{C. Cheng et al.}
%
\institute{Image X Institute, The University of Sydney, Sydney, Australia 
\email{chen.cheng@sydney.edu.au}}
\maketitle              
\begin{abstract}
To enable accurate and rapid photon control point and proton beamlet dose calculation in the DoseRAD2026 challenge\footnote{\url{https://doserad2026.grand-challenge.org/}}, we present BEAM3R, a dose estimation framework operating in beam's-eye-view (BEV). Our core innovation combines a Mamba-3 state-space depth-sequence core with physics-based transport conditioning to model long-range depth transport without expensive 3D convolutions. BEAM3R shares a 2D CNN encoder-decoder architecture for photon and proton dose tasks, processing per-plane BEV slices. Proton beamlets are conditioned on water equivalent thickness and remaining range, encoding the parameters determining Bragg peak position. Photon models use a bidirectional Mamba-3 core to capture dose contributions from materials downstream of the calculation point, while the proton model uses a forward core with learned energy-prefix tokens and a Bragg-peak refinement module. To reduce interpolation artifacts and support high spatial resolution, we introduce axial grid alignment of BEV lattices with CT slices and an implicit super-resolution representation arranged as channels, evaluated by a differentiable Triton-accelerated resampler that reconstructs packed cubic B-spline coefficients directly in CT space. For MRI-based tasks, synthetic CTs (sCT) are generated using a conditional GAN with a SwinUNETR generator and a PatchGAN discriminator. On the preliminary DoseRAD2026 test set, CT-to-photon and CT-to-proton models achieved 1\%/1 mm local gamma pass rates of 96.8\% and 96.0\%, with stratified plan-level MAEs of 0.0041 and 0.0079. Substituting sCT reduced gamma pass rates to 89.7\% for photon and 75.4\% proton plan level doses, with stratified plan-level MAEs of 0.0093 and 0.0336. Standardised runtimes were 23.4 s and 18.4 s for CT-to-photon and CT-to-proton prediction, increasing to 39.7 s and 42.8 s for the corresponding MRI-based pipelines.
\keywords{deep learning  \and proton dose calculation \and photon dose calculation \and synthetic CT \and Mamba-3.}
\end{abstract}
\section{Introduction}\label{sec:intro}

Monte Carlo (MC) simulation is the gold standard for radiotherapy dose calculation \cite{Rogers_2006}, but its computational cost limits its use in time-sensitive applications such as online adaptive radiotherapy \cite{Lim_Reinders_2017} and dose-guided radiotherapy \cite{Keall_2025}. The DoseRAD2026 Grand Challenge benchmarks rapid photon and proton beam-level dose calculation from CT and MRI. Beam-level modelling enables learning of beam-to-anatomy dose relationships, supporting flexible plan optimisation and dose accumulation workflows \cite{Kontaxis_2020}. 

Beam's-eye-view (BEV) slice-sequence models represent dose transport along a common beam-depth axis, reducing the need to encode treatment geometry explicitly while supporting efficient beam-level prediction \cite{neishabouri2021long,pastor2023sub,xiao2026multi}. Following this approach, we formulate 3D dose calculation using a 2D CNN encoder, a depth-sequence core, and a 2D CNN decoder.

We build on the BEV CNN-sequence architecture of Xiao et al. \cite{xiao2026multi} and introduce targeted modifications to improve beam coverage, BEV-to-CT resampling accuracy, model capacity, and downstream depth context. Photon and proton models share this framework but use task-specific sequence cores and inputs, including proton transport conditioning. For the MRI tasks, a SwinUNETR-based \cite{hatamizadeh2021swin} conditional generative adversarial network first generates a synthetic CT (sCT), which is passed to the corresponding CT-to-dose model. We apply these methods to all four challenge tasks: CT- and MRI-based photon dose calculation and CT- and MRI-based proton dose calculation.

\section{Methods}\label{sec:methods}
\subsection{Data}\label{subsec:data}
Separate models were trained for each task using only the DoseRAD2026 training dataset \cite{xiao-doserad-dataset}. Table~\ref{tab:splits} summarises the patient-level splits used by the tasks. Patient-level splitting ensured that no control point or beamlet from a validation patient appeared in training. Hyperparameters and checkpoints were selected using the held-out validation cohorts. 

\begin{table}[h!]
\centering
\caption{Patient-level training and validation splits. Patient 1ABB102 was excluded for photon model as it had outlier statistics in dose fluence. The abdomen split was unused for the final submitted proton model as the was only trained on thorax patients.}\label{tab:splits}
\footnotesize
\begin{tabular}{|p{2.5cm}|p{1.8cm}|>{\raggedleft\arraybackslash}p{1.5cm}|p{4.0cm}|>{\raggedleft\arraybackslash}p{1.2cm}|}
\hline
Task/modality & Cohort & Training patients & Validation patients & Split seed\\
\hline
Photon & Abdomen & 30 & \texttt{1ABB041}, \texttt{1ABB070}, \texttt{1ABB110}, \texttt{1ABB115}, \texttt{1ABB155} & 333\\
Photon & Thorax  & 36 & \texttt{1THB017}, \texttt{1THB023}, \texttt{1THB121} & 333\\
Proton & Abdomen & 34 & \texttt{1ABB030}, \texttt{1ABB083} & 42\\
Proton & Thorax  & 37 & \texttt{1THB095}, \texttt{1THB211} & 42\\
\hline
\end{tabular}
\end{table}
\begin{figure}[t!]
\centering
\includegraphics[width=0.88\textwidth]{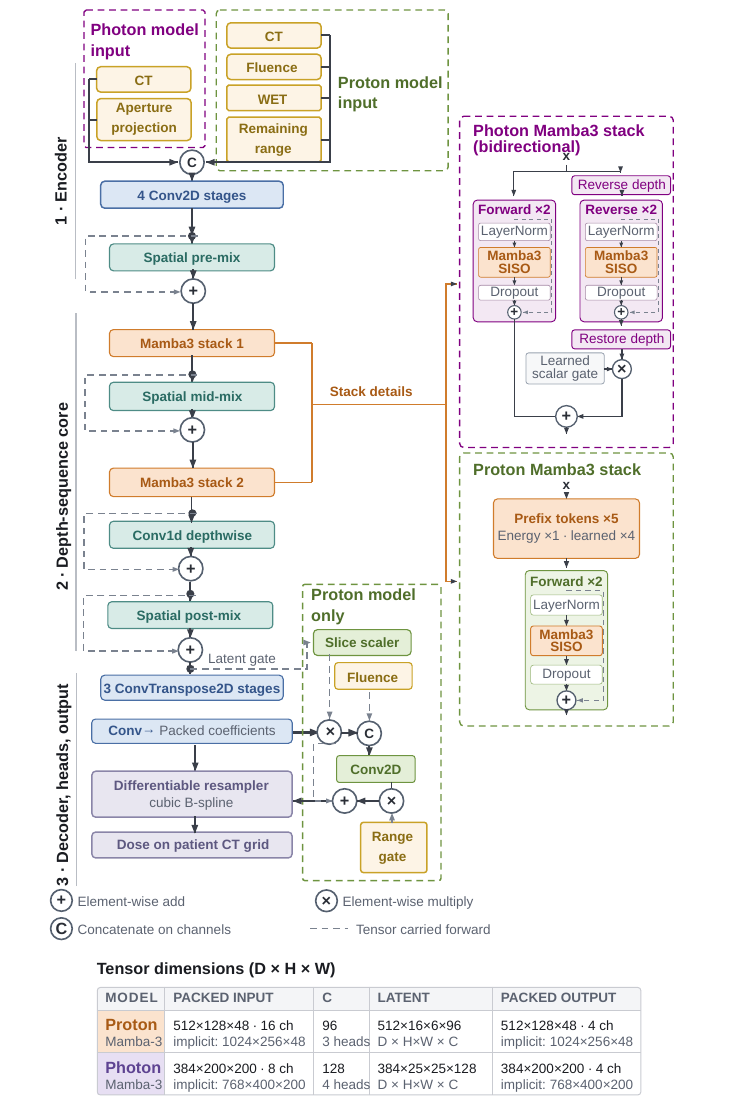}
\caption{Shared BEV architecture for photon and proton dose prediction. Green dashed boxes mark proton-only components  and the magenta dashed box shows the photon-only components. The table summarises the tensor dimensions as $(D,H,W)$, where $D$ is the beam depth (sequence axis), $H$ is the lateral in-plane axis which rotates with the gantry, and $W$ is parallel to the patient's superior-inferior axis. }
\label{fig:proton-arch}
\end{figure}

\subsection{Model}\label{subsec:model}
\textbf{Architecture overview.} Figure~\ref{fig:proton-arch} shows the architecture design used for our dose prediction models. Our models build on the BEV CNN-sequence architecture of Xiao et al. \cite{xiao2026multi}, retaining its strided encoder and decoder, residual depthwise-pointwise spatial mixing, and residual sequence blocks. The photon model predicts the dose from one control point, whereas the proton model predicts the dose from one beamlet. Complementary measures reduce BEV-to-CT resampling error: at input, we use an implicit super-resolution representation in combination with lattice snapping, which aligns one of the BEV lattice samples along one direction with CT lattice samples. Packed spline coefficients provide a finer implicit output field without materialising a dense BEV volume. A per-plane CNN encoder extracts spatial features from BEV slices, a sequence core propagates information along beam depth, and a per-plane decoder predicts packed cubic B-spline coefficients. A differentiable resampler evaluates the resulting continuous dose field on the patient CT grid. 

\textbf{BEV geometry and inputs.} BEV coordinates align dose transport with a common depth axis and avoid explicitly passing gantry angle, isocentre, and beam direction information as inputs into the neural network by using these geometrical parameters to directly perform sequence modelling along the beam direction \cite{neishabouri2021long,pastor2023sub,xiao2026multi}. The photon BEV depth was increased from the 51.2~cm extent used by Xiao et al. \cite{xiao2026multi} to 76.8~cm to avoid truncating long thoracic and abdominal beam paths. Photon inputs comprised of the normalised CT and projected MLC aperture, while proton inputs consisted of the normalised CT, Gaussian spot fluence, cumulative water-equivalent thickness (WET), and remaining range. 

\textbf{BEV lattice sampling.} BEV sample locations along the S--I axis were chosen to coincide with the native CT slice centres so that no interpolation is required along that direction. Fine-lattice sampling packed into separate channels reduced discretisation error when creating the BEV lattice in the remaining depth and lateral in-plane directions. The coarse photon lattice was $384\times200\times 200$ samples with $2$ mm isotropic spacing, covering $768\times400\times 400$ mm$^{3}$ along the depth, lateral and S–I axes. The coarse proton lattice was $512\times128\times48$ samples with $(1\times1\times3)$ mm sampling, covering $512 \times 128 \times 144$ mm$^{3}$ along the depth, lateral and S-I axes. However, the CT and derived inputs were sampled on a two-fold finer lattice (1 mm photon, 0.5 mm proton), giving an effective input sampling of $768 \times 400 \times 200$ and $1024 \times 256 \times 48$ respectively for photon and proton. The four sub-pixel phases were packed into input channels, so that the encoder grid and Mamba-3 depth sequence length were unchanged by the finer sampling.

\textbf{Proton model physics conditioning.} Our key innovation is that we supply derived transport quantities to the network. To encode the physical determinant of Bragg-peak position directly, the proton model was conditioned on cumulative WET and remaining range along each ray. WET was calculated from an energy-dependent relative stopping-power estimate as
\begin{equation}
\mathrm{WET}(d) = \int_{0}^{d} \mathrm{RSP}(s)\,\mathrm{d}s .
\end{equation}
 where the relative stopping-power ratio (RSP) was computed per voxel from the Geant4 HU-to-material calibration as the product of the relative electron density and the Bethe stopping-number ratio to water at the beamlet energy. The remaining range 
\begin{equation}
R_{\mathrm{rem}}(d) = R(E) - \mathrm{WET}(d).
\end{equation}
was calculated using the CSDA range approximation of Bortfeld \cite{bortfeld1997} over the dataset energy range of 31.729--200.797~MeV spanned by the dataset.
\begin{equation}
R(E) = 0.022\,E^{1.77}\ \mathrm{mm},
\end{equation}
The input also included a Gaussian fluence map parameterised by the beamlet spot width. The spot width $\sigma$ is taken from the beamlet metadata where present and otherwise from the machine's energy--spot table, and enters as $\exp[-(y^{2}+z^{2})/(2\sigma^{2})]$ about the central ray. Gaussian tails falling outside the cuboid are truncated to zero without renormalisation; with a maximum $\sigma$ of 7.99\,mm against a 128\,mm transverse extent. These continuous transport quantities directly encode the expected Bragg-peak location and were computed for each beamlet.

\textbf{Encoder.} The encoder comprises of four convolutional layers, with three strided layers providing eightfold in-plane downsampling. It maps the inputs to a latent feature grid, using LeakyReLU activations after each convolution. A residual spatial mixing block, consisting of a depthwise $7\times7$ convolution followed by pointwise $1\times1$ channel mixing, refines the encoded features. The per-slice latent representation is then rearranged into depth-ordered sequences so that each spatial position forms an independent feature sequence ordered along beam depth.


\textbf{Sequence Core.} The recurrent core mixes information along depth in the coarse grid latent space. Both proton and photon dose models use single-input-single-output (SISO) Mamba-3 blocks \cite{lahoti2026mamba}, chosen for accuracy without the slower training throughput of the multi-input, multi-output (MIMO) variant. Mamba-3 uses a complex-valued exponential--trapezoidal state-space recurrence, which provides stronger state tracking and associative recall than the previous iteration (Mamba-2), keeps a fixed-size recurrent state and linear-time cost, and is deployed with memory-efficient chunked training and Triton kernels for GPU acceleration. Similar to Xiao et al \cite{xiao2026multi}, we adopt depthwise spatial convolutions and pointwise channel mixing in residual spatial-mixing blocks to aggregate spatial neighbourhood information and fuse information between channels. These spatial-mixing blocks are applied before, between, and after the two sequence-core passes. An additional residual depthwise 1D convolution supplies local, non-causal depth context. Motivated by previous embeddings of the beam energy to condition proton dose predictions \cite{li2025neural,li2025proof,pastor2022millisecond}, the proton model additionally includes a learned per-energy embedding which is prepended as a prefix token to the Mamba temporal sequence. For photon dose prediction, a second Mamba-3 branch scans the reversed depth sequence and is fused with the forward branch through a learnable, zero-initialised gate, enabling the model to predict dose contributions from material downstream
of it, principally scatter, which a causal core cannot represent.

\textbf{Decoder and CT space reconstruction.} 
The latent features from depth-sequence modelling are reshaped into depth-indexed feature grids and decoded slice-wise in 3 transposed-convolution stages with LeakyReLU or ReLU activation, followed by a $1\times1$ convolution producing four packed coefficient channels using sub-pixel upsampling \cite{shi2016real}. The head predicts four cubic B-spline coefficient channels \cite{witte2024deep}, representing a two-fold finer dose field along depth and height without explicitly materialising a high-resolution BEV volume. The four packed output channels correspond to the four sub-pixel phases of a cubic B-spline coefficient lattice at $2\times$ finer resolution along BEV depth and the in-plane lateral direction. For the proton model only, the coefficients are firstly refined by a residual convolutional head, whereas the photon coefficients are passed directly to the resampler. The residual convolutional head consists of a per-plane 3 × 3 convolution, a SiLU activation and a 1 × 1 convolution; it takes the packed coefficients and packed fluence and predicts an additive correction in the coefficient domain. The correction is gated by the packed remaining range, which concentrates it on the Bragg peak, and the final convolution is zero-initialised so the head is the identity at initialisation. 

The patient-$z$-aligned axis remains on the native CT slice grid and therefore does not require cubic interpolation. For a CT voxel centre $\mathbf r_j$, let 
\begin{equation}
(u_j,v_j,w_j)=\mathcal{T}^{-1}\mathbf r_j ,
\end{equation}
where $\mathcal{T}^{-1}$ maps patient CT coordinates into BEV coordinates. The predicted dose is evaluated as
\begin{equation}
\hat y(\mathbf r_j)
=
\sum_{n_u,n_v}
c_{n_u,n_v,k(w_j)}
\,
\beta_3\!\left(
\frac{u_j-o_u}{h_u}-n_u
\right)
\beta_3\!\left(
\frac{v_j-o_v}{h_v}-n_v
\right),
\end{equation}
where $c_{n_u,n_v,k}$ are the predicted spline coefficients, $\beta_3$ is the cubic B-spline basis, $h_u$ and $h_v$ are the fine-lattice spacings, $(o_u,o_v)$ are the corresponding lattice origins, and $k(w_j)$ denotes the CT-slice index associated with the aligned patient-$z$ coordinate. Thus, cubic interpolation is performed only in the two non-aligned BEV dimensions, while the aligned axis is evaluated directly at the corresponding CT slice. The continuous spline field is evaluated at the native CT voxel centres by a differentiable, custom Triton resampler, so the training loss is applied after the BEV-to-CT transformation rather than to an intermediate coarse BEV dose volume. The model is therefore optimised directly against dose in patient CT coordinates, the domain in which the prediction is evaluated and used, rather than against an intermediate BEV representation.

\textbf{Synthetic CT Generation.}\label{subsec:sct_generation}
For magnetic resonance imaging (MRI) tasks, MRI was converted to sCT using a cGAN. We used a SwinUNETR generator \cite{hatamizadeh2021swin} with feature size 48 trained with a body-masked image reconstruction loss and an adversarial term from a 3D PatchGAN discriminator \cite{isola2017image}. The generator operated on $64\times160\times160$ patches. MRI volumes were standardised per subject, CT volumes using training-population statistics. The resulting sCT was passed to the corresponding CT-based dose model.
\textbf{Parameters and initialisation.} The photon and proton dose models contain 1.77\,M and 0.73\,M trainable parameters, respectively; the sCT cGAN contains 75.41\,M parameters (72.76\,M generator; 2.64\,M discriminator). Convolutional and linear layers used PyTorch's default Kaiming-uniform initialisation. The photon model was warm-started from a unidirectional Mamba checkpoint, with its backward-fusion gate zero-initialised, while the range-conditioned proton model was warm-started from a model trained without the Bragg residual head. The initial dose checkpoints and both sCT models were trained from random initialisation. No external pretrained weights were used.


\textbf{Implementation.} Models were implemented in PyTorch using mamba-ssm 2.3.2 for the Mamba-3 cores and Triton 3.5.0 for custom packed-spline evaluation and BEV-to-CT resampling kernels. SimpleITK handled medical-image I/O and spatial transformations, while MONAI 1.6.0 provided data-loading components and the SwinUNETR sCT generator. Source code, complete package versions, and model configurations will be made available at \url{https://github.com/Image-X-Institute/DoseRAD2026\_BEAM3R}.

\textbf{Inference Optimisation.} The sequence core used the chunked scan and Triton kernels provided by mamba-ssm, and all models were evaluated in bfloat16 precision with fixed input dimensions. After checkpoint loading, inference-only kernel fusion and memory optimisations were applied, including fused Mamba-3 projection and normalisation, removal of training-only buffers, and fused decoder operations. These changes reduced memory traffic and kernel launches without altering model parameters or predicted dose. All Triton and CuPy kernels were compiled during a warm-up pass at container start-up. Dose was evaluated only within the CT bounding box intersecting each beam, while patient CT data and spline coefficients were cached across dose stacks. Output transfer, MetaImage compression, and file writing were performed asynchronously to overlap input-output operations with GPU execution. The Mamba-3 chunk size (16) and compression level (deflate level 1) were selected empirically to minimise end-to-end runtime.

\subsection{Training}\label{subsec:training}
\textbf{Dose-model training.} CT values were clipped to $[-1024,3071]$ and rescaled to $[0,1]$. Reference doses were normalised using a scalar estimated from the training patients and restored to physical units during inference. No explicit data augmentation was applied to the dose models. Beam-specific BEV inputs were generated on-the-fly from the treatment geometry.

We used a staged learning approach for our dose prediction models to accelerate model convergence, with the added benefit of being able to learn primary transport physics before learning secondary residual corrections. Photon training comprised of three stages. First, a unidirectional model was trained from random initialisation on the thorax cohort for 200 epochs using an initial learning rate of $2.5\times10^{-4}$. This checkpoint was then fine-tuned on the combined thorax and abdomen training cohorts for 60 epochs using an initial learning rate of $1\times10^{-4}$. Finally, the reverse-depth Mamba branch was added and the resulting bidirectional model was fine-tuned for a further 100 epochs, also with an initial learning rate of $1\times10^{-4}$. The proton model was trained only on the thorax training cohort from random initialisation for 200 epochs without the Bragg residual head. The residual head was then added and the complete model fine-tuned for 200 epochs. Both proton stages used an initial learning rate of $1\times10^{-4}$. 

Both dose models minimised mean squared error (MSE) between the predicted and reference dose distributions and included a masked mean absolute error (MAE) term to emphasise clinically relevant dose regions. Losses were evaluated on the reconstructed patient CT grid. The mask 
\begin{equation} M = \mathbf{1}\!\left(y \geq 0.1\,\max(y)\right) \end{equation}
selects voxels receiving at least 10\% of the maximum reference dose.

The photon models were optimised using:
\begin{equation} \mathcal{L}_{\mathrm{photon}} = \mathrm{MSE}(\hat y,y) + \lambda_{\mathrm{MAE}} \frac{\sum_i M_i |\hat y_i-y_i|} {\sum_i M_i}, \end{equation}

with $\lambda_{\mathrm{MAE}} = 2\times10^{-4}$. The proton models used the same objective together with an additional non-negativity penalty applied within the valid sampling region:
\begin{equation} \mathcal{L}_{\mathrm{proton}} = \mathrm{MSE}(\hat y,y) + \lambda_{\mathrm{MAE}} \frac{\sum_i M_i |\hat y_i-y_i|} {\sum_i M_i} + \lambda_{\mathrm{valid}} \frac{\sum_i V_i\,\max(0,-\hat y_i)^2} {\sum_i V_i}, \end{equation}

where $\lambda_{\mathrm{MAE}} = 1\times10^{-4}$, $\lambda_{\mathrm{valid}} = 1\times10^{-2}$, and $V$ denotes the valid sampling region. The final term penalises physically implausible negative dose predictions.

All dose models were trained using Adam with batch size 4, no weight decay, and global gradient-norm clipping at 1.0, with 8 data-loader workers for the photon runs and 4 for the proton runs. The learning rate was halved when validation loss did not improve for 10 photon or 15 proton epochs. Training used bfloat16 mixed precision.

\textbf{Synthetic CT training.} Following the SynthRAD2025 KoalAI (winning MRI to CT model submission) preprocessing approach \cite{rogowski2026generating,thummerer2025synthrad2025}, where the MRIs were standardised using a subject z-score while the CTs were standardised using a population mean and standard deviation, enabling the recovery of HU by inverse normalisation based on these population-level statistics. A body mask restricted optimisation to anatomically relevant voxels. The sCT cGAN was trained on $64\times160\times160$ patches for up to 1000 epochs. Data augmentation comprised of random cropping and flipping, intensity and contrast perturbations, and additive Gaussian noise. The sCT model was optimised using: 
\begin{equation}
\mathrm{\mathcal{L}_{G,total}} = \mathrm{\lambda_{G,image} \cdot \mathcal{L}_{G,image} + \mathcal{L}_{G,adv},}
\end{equation}
where $\mathrm{\lambda_{G,image}}$ is the generator image content weighting term (set to 100 for the proton tasks and 50 for the photon tasks), $\mathrm{L_{G,image}}$ is the generator image content loss term (MSE), and $\mathrm{L_{G,adv}}$ is the adversarial loss term (MSE, based on least-squares GAN \cite{mao2017least}). The generator and discriminator were optimised using AdamW \cite{loshchilov2017decoupled} and Adam \cite{kingma2014adam}, respectively. The checkpoint with the lowest validation HU MAE was retained. 

\textbf{Hyperparameter and model selection.} Hyperparameters were selected through targeted experiments on the held-out patients described in Section~\ref{subsec:data}. Within each training stage, the checkpoint with the lowest validation loss was retained. Candidate model configurations were compared using the official beam-level metrics on the held-out validation set. No model ensembling was used, which makes for efficient runtime.


\subsection{Evaluation}\label{subsec:evaluation}

During training, MSE and masked MAE were monitored on the held-out validation patients. Beam-level performance was evaluated using the challenge-provided code for: (i) masked MAE, calculated over voxels receiving at least 10\% of the reference beam maximum and normalised by that maximum; and (ii) integrated depth-dose (IDD) distance, defined as the normalised root-mean-square difference between predicted and reference IDD curves along the beam direction.

Beam-level results were averaged within each patient and summarised across patients using the mean and standard deviation. Individual patient values are also reported. Given the small validation cohorts, no confidence intervals or formal significance tests were calculated.

Packaged models were additionally evaluated on the hidden test set through the DoseRAD2026 Grand Challenge platform. Official plan-level metrics were: (i) stratified MAE, averaged across low-dose (10--30\%), intermediate-dose (30--80\%), and high-dose ($\geq80\%$) regions; (ii) the 3D local gamma pass rate at 1\%/1\,mm; and (iii) a DVH-based clinical score combining relative errors in PTV $D_{98\%}$ and $V_{95\%}$ and in $D_{2\%}$ and mean dose for the three organs-at-risk closest to the target.

Runtime was estimated by the organisers using a non-negative linear model of fixed, per-image, and per-dose-map costs. The fitted model was evaluated for representative cases comprising one image and either 181 photon control points or 500 proton beamlets.

No formal model-explainability method was applied. Depth-resolved IDD curves and spatial dose-error maps were instead inspected as qualitative diagnostics of systematic dose errors, including dose mismatch and penumbra misalignment.
\section{Results}\label{sec:results}

\textbf{Challenge performance} Table~\ref{tab:results-best} summarises performance on the DoseRAD2026 preliminary test set. The CT-based photon and proton models achieved similar beam-level accuracy, with masked MAE of 0.0084 for both and local gamma pass rates of 96.8\% and 96.0\%, respectively. The corresponding MRI-based tasks produced higher errors, particularly for proton dose prediction: relative to CT-to-proton prediction, MRI-to-proton masked MAE increased from 0.0084 to 0.0326, while the gamma pass rate decreased from 96.0\% to 75.4\%. The substantially larger degradation for MRI-to-proton prediction highlights sCT accuracy as a major limitation for MRI-based proton dose calculation. Runtime on an AWS instance with A10G GPU, 4 vCPUs and 16GB RAM on Grand Challenge was around 22-23s for the CT to photon/proton tasks and increased to 39-43s attributed to time taken to perform sCT inference. Local runtime for inference running on a NVIDIA RTX A6000 was 16.7s for 180 photon beams and 8.176s for 300 proton beamlets. 

\begin{table}[htbp]
\centering
\caption{Evaluation metrics for CT/MRI to photon/proton dose calculated on the preliminary test set for the DoseRAD2026 challenge. MAE: beam-level mean absolute dose error; IDD Dist.: normalised integrated depth-dose curve distance; Plan MAE: stratified plan-level MAE; $\gamma$ Index: 3D local gamma pass rate (\%) at 1\%/1\,mm; DVH Score: DVH-based clinical score.}
\label{tab:results-best}
\small
\setlength{\tabcolsep}{2.5pt}
\begin{tabular}{c*{6}{c}}
\toprule
Task & MAE & IDD Dist. & Plan MAE & $\gamma$ Index & DVH Score & Runtime (s) \\
\midrule
CT to photon & \shortstack{0.0084\\{\scriptsize$\pm$0.0022}} & \shortstack{0.0069\\{\scriptsize$\pm$0.0105}} & \shortstack{0.0041\\{\scriptsize$\pm$0.0019}} & \shortstack{96.7569\\{\scriptsize$\pm$3.0172}} & \shortstack{0.2173\\{\scriptsize$\pm$0.0663}} & 23.4493 \\
MRI to photon & \shortstack{0.0139\\{\scriptsize$\pm$0.0064}} & \shortstack{0.0097\\{\scriptsize$\pm$0.0093}} & \shortstack{0.0093\\{\scriptsize$\pm$0.0056}} & \shortstack{89.7046\\{\scriptsize$\pm$7.2553}} & \shortstack{0.9051\\{\scriptsize$\pm$0.1782}} & 39.6613 \\
CT to proton & \shortstack{0.0084\\{\scriptsize$\pm$0.0039}} & \shortstack{0.0058\\{\scriptsize$\pm$0.0081}} & \shortstack{0.0079\\{\scriptsize$\pm$0.0039}} & \shortstack{95.9532\\{\scriptsize$\pm$3.6125}} & \shortstack{0.5018\\{\scriptsize$\pm$0.3073}} & 18.3807 \\
MRI to proton & \shortstack{0.0326\\{\scriptsize$\pm$0.0216}} & \shortstack{0.0287\\{\scriptsize$\pm$0.0208}} & \shortstack{0.0336\\{\scriptsize$\pm$0.0226}} & \shortstack{75.4368\\{\scriptsize$\pm$18.9888}} & \shortstack{9.5124\\{\scriptsize$\pm$9.5922}} & 42.8198 \\
\bottomrule
\end{tabular}
\end{table}

\textbf{Ablation experiments} Table~\ref{tab:ablation} summarises pairwise development experiments assessing BEV extent, model capacity, sequence core, depth directionality, CT-space training, grid snapping, and super-sampling. Model capacity is set by a single width multiplier which scales every convolutional channel count and sequence-core feature width. For each comparison, the lowest-validation-loss checkpoint within the epoch range shared by both runs was used. Because experiments were conducted at different development stages, the cohort and secondary metric vary by comparison; results should therefore be interpreted within, rather than across, pairs.

\begin{table}[!t]
\centering
\caption{Experiments motivating design decisions final model. Each pair varies the
stated factor; the epoch is the lowest-validation-loss epoch within the range common to both
runs. Metric sets differ by experiment generation:
$^{a}$ BEV-space masked MAE and BEV IDD distance;
$^{b}$ CT-space masked MAE and round-trip BEV IDD distance;
$^{c}$ CT-space masked MAE and true-gantry beam IDD distance;
$^{d}$ CT-space masked MAE and CT-space MSE ($\times10^{-6}$), the two runs sharing no IDD
variant for the reason given in the text.
Rows marked $\dagger$ vary one further factor and $\ddagger$ several (see text). Lower is
better throughout. SR: super-resolution.}
\label{tab:ablation}
\small
\setlength{\tabcolsep}{4pt}
\begin{tabular}{llccc}
\toprule
Design decision & Variant & Epoch & MAE & IDD \\
\midrule
\multicolumn{5}{l}{\textit{Thorax cohort, 36 training patients}} \\
BEV depth extent$^{b}$
  & 256 planes            & 122 & 0.0117 & 0.0120 \\
  & \textbf{384 planes}   & 124 & \textbf{0.0107} & \textbf{0.0084} \\[2pt]
Model capacity$^{b}$
  & $1.0\times$           & 94  & 0.0116 & 0.0076 \\
  & $\mathbf{2.0\times}$  & 96  & \textbf{0.0098} & \textbf{0.0070} \\[2pt]
Sequence core$^{b,c}$ $\ddagger$
  & mLSTM                 & 115 & 0.0104 & -- \\
  & \textbf{Mamba-3}      & 138 & \textbf{0.0090} & -- \\
\midrule
\multicolumn{5}{l}{\textit{Merged cohort, warm-started from the unidirectional run}} \\
Depth directionality$^{c}$
  & unidirectional        & 72  & 0.0075 & 0.0043 \\
  & \textbf{bidirectional}& 75  & \textbf{0.0071} & \textbf{0.0034} \\
\midrule
\multicolumn{5}{l}{\textit{Pilot cohort, 10 training patients, thorax and abdomen}} \\
CT-space training
  & BEV-space training       & 100  & 0.0195 & -- \\
  & \textbf{CT-space + SR}   & 100 & \textbf{0.0132} & -- \\[2pt]
Axial grid snapping$^{a}$ $\dagger$
  & off                   & 60  & 0.0150 & 0.0135 \\
  & \textbf{on}           & 65  & \textbf{0.0140} & \textbf{0.0125} \\[2pt]
Super-sampled inputs$^{d}$
  & off                   & 68  & 0.0141 & -- \\
  & \textbf{on}           & 66  & \textbf{0.0138} & -- \\
\bottomrule
\end{tabular}
\end{table}
All pairwise comparisons favoured the adopted configuration. On the thorax cohort, increasing BEV depth from 256 to 384 planes reduced masked MAE by 8.5\% and IDD distance by 30.0\%, while doubling model capacity reduced masked MAE by 15.5\%. Adding the reverse-depth Mamba branch reduced masked MAE by 5.3\% and the IDD distance by 20.9\%. Pilot experiments also favoured CT-space training with axial grid snapping, implicit super-resolution inputs and outputs to reduce interpolation error between BEV and CT-space. Several pairs differed in additional settings, so these results indicate associations rather than isolated component effects.

\begin{figure}[bt!]
    \centering
    \includegraphics[width=0.9\linewidth]{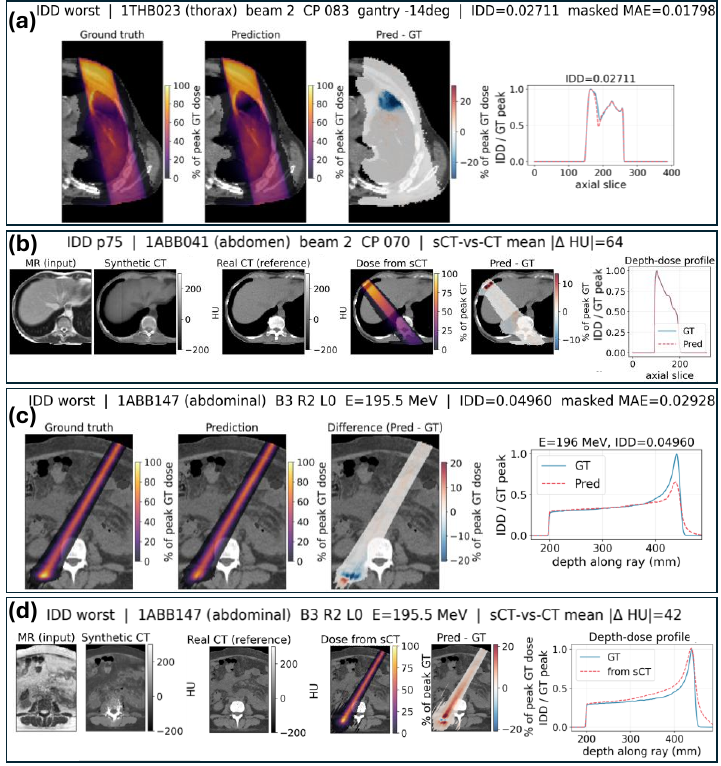}
    \caption{Failure modes in photon and proton dose estimates taken from the held out validation set. (a) and (c) show a comparison of ground truth, prediction, difference dose and IDD curve prediction for a CT input to photon beam prediction and a proton beamlet prediction respectively. (b) and (d) show a comparison of the MR input, sCT, real CT, prediction and  dose difference for a photon beam and a proton beamlet respectively.}
    \label{fig:results-plots}
\end{figure}

\textbf{Qualitative error analysis} Figure~\ref{fig:results-plots} compares predicted and reference dose for photon beams and proton beamlets with high IDD error and were selected to illustrate failure modes. The prediction reproduced the overall beam trajectory while the largest discrepancies occurred at peak dose deposition. The corresponding IDD curves showed the mismatch between ground truth and predicted doses and a range shift for the proton case which is consistent with the spatial difference map. For the MRI-based tasks, (b) shows visible striation artifacts in the sCT. In both (b) and (d) inaccurate soft tissue and bone HU coincided with local dose errors. 
\section{Discussion}\label{sec:discussion}
\textbf{Performance across input modalities.} The CT-based photon and proton models achieved similar performance, with local gamma pass rates of approximately 96\% at 1\%/1\,mm and normalised masked MAE below 0.9\% of the reference beam maximum. Performance decreased when planning CT was replaced by sCT, with a substantially larger reduction for proton than photon prediction. Relative to the corresponding CT task, the MRI-to-photon gamma pass rate decreased by 7.3\%, whereas the MRI-to-proton rate decreased by 21.4\%.

The greater proton sensitivity is consistent with the dependence of proton range on integrated stopping power. HU errors along a beam path can shift the predicted Bragg peak, whereas comparable errors generally perturb photon attenuation without producing the same depth displacement. In addition, sCT error affects the proton model both through the image and through the WET and remaining-range volumes derived from it. The qualitative examples show spatial coincidence between sCT inaccuracies and dose discrepancies, but do not establish causality.

\textbf{Architectural design choices.} The development experiments favoured greater model representational capacity: BEV extent, increased model width, and bidirectional depth modelling. Grid alignment along the S--I direction and CT-space training with super-sampling over the remaining spatial dimensions were favoured in the development experiments. Geometrically, S--I alignment removes inter-slice interpolation during CT-to-BEV sampling, while packed fine-grid sampling reduces discretisation in the remaining coordinates and the continuous spline representation permits direct evaluation of the predicted dose at CT voxel centres rather than interpolation of a coarse BEV dose volume. Negligible impact on inference time was achieved through a channel-rather-than-spatial representation of the super-sampled grid alongside dedicated Triton kernels for the resampling. The largest observed accuracy improvements were associated with this super-sampling, BEV extent, model width, and bidirectional depth processing. In particular, the reverse-depth Mamba branch improved IDD distance more than masked MAE, consistent with its intended role in modelling depth-dependent effects that are unavailable to a strictly forward scan. However, the experiments used different cohorts and metric implementations, and several comparisons varied additional settings; their effect sizes are therefore not directly comparable or attributable to single components.

\textbf{Limitations.} The validation cohorts contained only two or three patients per task, preventing reliable confidence intervals or formal patient-level significance testing. While the development experiments provided directional insights motivating our architectural choices, they were conducted iteratively under strict challenge deadlines and computation constraints and the design of the ablation study could be improved to strictly isolate a single variable and use consistent evaluation metric definitions. The temporal convolution, energy prefix, slice scaler, and Bragg residual head were not independently ablated. Due to compute and time constraints, proton fine-tuning was not completed on the full training cohort, potentially limiting anatomical generalisability. The sCT model performance may also depend on scanner and acquisition characteristics, and the simplified RSP and range approximations may not generalise beyond the energies and anatomy represented in the challenge dataset. 

\section{Author Contributions}\label{sec:contributions}

D.E.J.W --- Writing -- review \& editing, Supervision \& Resources

E.H. --- Writing -- review \& editing, Supervision

C.C. -- Conceptualisation, Methodology, Software, Validation, Formal Analysis, Investigation, Data Curation, Writing -- Original Draft, Visualisation

M.F. -- Conceptualisation, Methodology, Software, Validation, Formal Analysis, Investigation, Writing -- Original Draft, Visualisation

J.G. -- Conceptualisation, Methodology, Software, Validation, Investigation, Writing -- Original Draft

\section{Other Information}\label{sec:other}

\begin{credits}
\subsubsection{\ackname} C.C., M.F. and J.G. are  supported by Australian Government Research
Training Program scholarships. D.E.J.W. received support from the Australian Government National Health and Medical Research Council Investigator Grant 2017140 and the University of Sydney Collaborative Research Equipment Grants and NHMRC Equipment Grants Scheme. E.H. and D.E.J.W. acknowledge funding from a Cancer Council NSW Project Grant (RG 25-01). 

\subsubsection{\discintname}
None to declare.
\end{credits}
%
%
\bibliographystyle{splncs04}
\bibliography{mybibliography}

\end{document}